\documentclass{llncs}
\usepackage{hyperref}
\usepackage{graphicx}
\usepackage{ragged2e}
\usepackage[final]{listings}
\usepackage[font=footnotesize]{subfig}
\usepackage{array}
\usepackage{multirow}
\usepackage{rotating}
\usepackage{listings}
\usepackage{mathptmx}

\newcommand{\ublas}{{\it uBLAS}}

\newcommand{\eigen}{{\it Eigen3}}

\graphicspath{{figures/}}

\begin{document}
\title{Best practices for HPM-assisted performance 
  engineering on modern multicore processors}
\author{Jan Treibig \and Georg Hager \and Gerhard Wellein}
\institute{Erlangen Regional Computing Center (RRZE)\\
Friedrich-Alexander-Universit\"at Erlangen-N\"urnberg\\
Martensstr. 1, 91058 Erlangen, Germany\\
\email{\{jan.treibig,georg.hager,gerhard.wellein\}@rrze.fau.de}}
\maketitle
\begin{abstract}
  Many tools and libraries employ hardware performance
  monitoring (HPM) on modern processors, and using this data for performance
  assessment and as a starting point for code optimizations is very
  popular. However, such data is only useful if it is interpreted with
  care, and if the right metrics are chosen for the right purpose.  We
  demonstrate the sensible use of hardware performance counters in the
  context of a structured performance engineering approach for
  applications in computational science. Typical performance patterns
  and their respective metric signatures are defined, and some of them 
  are illustrated
  using case studies. Although these generic concepts do
  not depend on specific tools or environments, we restrict ourselves
  to modern x86-based multicore processors and use the likwid-perfctr
  tool under the Linux OS.
\end{abstract}
\section{Introduction and related work}\label{sec:intro}

\begin{figure}[tbp]
    \centering\includegraphics*[width=0.8\linewidth]{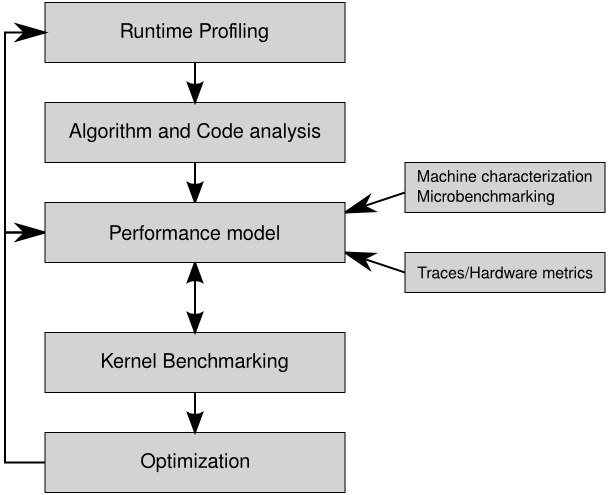}
    \caption{\label{fig:perfeng} Structured performance engineering process. The 
    HPM results come in via traces and hardware metrics, but machine
    parameters and microbenchmarking are of equal importance. A performance
    model for each core loop is constructed from this data and successively
    refined and adapted during the benchmarking and optimization process.}
\end{figure}

Hardware performance monitoring (HPM) is regarded as a state of the art
advanced tool to guide code optimizations. While there are countless
publications about HPM-based optimization efforts, a structured method
for using hardware events is often missing. 
even from the same vendor. 
One exception is the use of cache miss events, which are
very popular since memory access is regarded to be a major
bottleneck on modern architectures. In fact, miss events 
are often seen as the most useful metrics in HPM. Many
optimization efforts solely focus on minimizing cache miss ratios
\cite{guenther06cache-aware}.  Another popular application of HPM is automatic
performance tuning via a runtime approach
\cite{DBLP:journals/thipeac/KlugOWT11,Chen03dynamictrace}. Recent
work attempts to apply statistical methods such as regression analysis to achieve
automatic application characterization based on HPM \cite{delaCruz20112146,4536278}. 

This paper will present a more holistic
view on how to use HPM in a sensible way. It suggests HPM as one
aspect embedded in a structured performance engineering approach (see
Fig.~\ref{fig:perfeng}).  The central idea is the iterative
development of a diagnostic performance model enforcing a better understanding
of the code properties and the hardware capabilities, leading to a deeper
understanding how a code interacts with a given architecture. HPM is one
important source of information to improve this understanding.
We want to stress that HPM is in many cases only meaningful if related to
other information like, e.g., microbenchmark results or static code analysis. In
the following we will concentrate on what role HPM can play in order to
identify performance properties and problems, and to
implement a structured optimization effort. 

While all these ideas are not new, we believe that the emphasis on
``patterns'' can help tackle problems that are not so easily
recognized using automatic tools that are too focused on HPM.
We concentrate here on the typical patterns and their identification.
The structured performance engineering approach will be published
in detail elsewhere.

\subsection{Hardware performance metrics}

Hardware performance counters are available in every modern
microprocessor design. They allow the measurement of many (sometimes
hundreds) of metrics that are related to the way code is executed on
the hardware. Although many of those metrics are unimportant for the
developer writing numerical simulation code, some of them can be very
useful in assessing resource utilization and general performance
properties. A large variety of tools exist, from basic
to advanced, that allow easy access to HPM data, and some of them 
even give optimization advice derived from the measurements. 

Fortunately, although there is considerable variation in the kinds of hardware
events that are available on different processors (even from the same
manufacturer), a rather small subset of them is sufficient to identify the most
prevalent performance problems in serial and parallel code. These are available
on all modern processor designs.  We call a specific combination of hardware
event counts and possible other sources of information a ``signature.''
Together with information about runtime performance behavior and code
properties, signatures indicate the presence of so-called ``performance
patterns,'' which help to assess the quality of code and, most importantly,
identify relevant bottlenecks to enable a structured approach to performance
optimizations.

This paper is organized as follows. In Sect.~\ref{sec:patterns} we introduce a
(non-exhaustive) list of performance patterns and the typical
metric signatures that go with them. Sect.~\ref{sec:cases}
then presents two case studies from different sectors of computational
science, and Sect.~\ref{sec:summary} gives a summary and outlook
to future work.

\subsection{likwid-perfctr}

Given sufficient experience, simple and lightweight tools are often 
adequate to accomplish the goals described above. Hence, we restrict ourselves to x86
architectures under the Linux OS and employ the likwid-perfctr tool from the
LIKWID toolsuite \cite{likwid-psti,likwidweb}. LIKWID\footnote{``Like I Knew
What I'm Doing''} is a collection of command line programs that facilitate
performance-oriented program development and production in x86 multicore
environments under Linux. The concept of event sets with connected derived
metrics, which is implemented in likwid-perfctr by means of performance groups,
fits well to the signature approach presented in this paper. We will not
go into details on how to employ likwid-perfctr, since other tools and frameworks
can do similar things.

\section{Performance patterns and event signatures}\label{sec:patterns}

The following performance patterns have been found to be most useful when
analyzing scientific application codes on multicore-based nodes.
Other application domains may have different issues, but the basic
principle could still be applied. The categorization is to some
extent arbitrary, and some patterns are frequently found together.
\begin{itemize}
    \item \textbf{Load imbalance}\\
	Load balancing issues are an impediment for parallel
	scalability, and hence performance, and they should be
	resolved first. 
	
\item \textbf{Bandwidth saturation}\\
        Whenever the bandwidth of a shared data path is exhausted,
	there is a natural limit to scalability. Most frequently
	this happens on the main memory interface or the 
	(usually shared) outer-level cache (OLC).

\item \textbf{Strided or erratic data access}\\
        Cache-based architectures require contiguous data accesses
        to make efficient use of bandwidth due to the cache line
        concept. Strided access (often caused by inappropriate data
        structures or badly ordered loop nests) is one of the most
        frequent causes for low data transfer efficiency (between cache
        levels and to/from memory).

\item \textbf{Bad instruction mix}\\ 
	Inefficient code execution due to an
    	instruction mix that is inadequate to solve the problem can be a
    	complex issue. It encompasses diverse effects such as general
    	purpose instruction overhead created by inefficient compiler code
    	(often occurs with C++), but also the degree of vectorization or
    	the use of expensive operations like divide and square root.
	
\item \textbf{Limited instruction throughput}\\
	There is always a limit for the number of instructions that can
	be executed per cycle (e.g., 4 or 6), independent of their
	types. Even if a code does not hit this limit, it could
	still suffer from a bottleneck in a specific execution 
	port (such as load or multiply). Finally, dependencies could 
	cause pipeline bubbles, which further diminish the throughput.
	This pattern is closely
	related to the bad instruction mix pattern, but tends to
	require different code optimization strategies.

%
\item \textbf{Microarchitectural anomalies}\\
	This is a very architecture-specific pattern which may have
	different manifestations depending on the type of CPU. 
	Typical examples are false store forward aliasing, unaligned
        data accesses or instruction code, and shortage of load/store
        buffers.

\item \textbf{Synchronization overhead}\\
        Barriers at the end of parallel loops or locks protecting
	shared resources may have a large performance impact if
	the workload between such synchronization points is too small.
	This pattern may also incur
	secondary effects like load imbalance or bad instruction mix
	(see above).

\item \textbf{False cache line sharing}\\
        Different threads accessing a cache line (and at least one
        of them modifying it) lead to frequent evictions and
        reloads, impacting performance a lot.   

\item \textbf{Bad page placement on ccNUMA}\\
        All modern multi-socket servers are of ccNUMA type. Memory-bound
	codes must implement proper page placement in order to profit
	from the bandwidth advantages that ccNUMA provides. The two main
	problems with bad page placement are nonlocal data access
	and bandwidth contention, with load imbalance as a possible
	secondary effect.

\end{itemize}
Each of those patterns can be mapped to one or more ``signatures,'' which
consist of a combination of performance behavior (scalability, 
sensitivity to problem size, etc.) and a particular pattern in raw
or derived hardware metrics. While the former is often independent
of the underlying architecture, the latter is very hardware-specific.
Ideally a given tool should provide
these event sets and derived metrics in a similar way on all supported
processor architectures. likwid-perfctr tries to support this by
``performance groups.'' In Table~\ref{tab:signatures} we give
a correspondence of each performance pattern with its signatures
in the performance behavior and to the relevant
anomalies in hardware metrics (together with the likwid-perfctr
performance group, if available). In some cases the signature
also involves information from other sources such as microbenchmarks or
static code analysis, since some HPM signatures may be easily 
misinterpreted. We deliberately do not give any general optimization
hints, since optimization is only possible trough a thorough code
review together with a suitable performance model. 


\begin{table}
    \centering\footnotesize
    \begin{tabular}{>{\RaggedRight}m{2.5cm}|>{\RaggedRight}m{3.5cm}|>{\RaggedRight\arraybackslash}m{5.7cm}}
    & \multicolumn{2}{|c}{\bfseries Signature} \\\cline{2-3}
        \bfseries Pattern & \bfseries Performance behavior & \bfseries {HPM (and likwid-perfctr group(s))}\\
        \hline
Load imbalance & Saturating speedup & Different count of instructions retired\footnotemark[2] or floating point operations among cores (\verb+FLOPS_DP+, \verb+FLOPS_SP+)\\\hline
OLC bandwidth saturation& Saturating speedup across cores in OLC group & OLC bandwidth comparable to peak bandwidth of a suitable microbenchmark (\verb+L3+)\\\hline
Memory bandwidth saturation & Saturating speedup across cores sharing a memory interface & Memory bandwidth comparable to peak bandwidth of a suitable microbenchmark (\verb+MEM+)\\\hline
Strided or erratic data access & Large discrepancy between simple bandwidth-based model and actual performance & Low bandwidth utilization despite LD/ST domination / Low cache hit ratios, frequent \mbox{evicts}/replacements (\verb+CACHE+, \verb+DATA+, \verb+MEM+)\\\hline
Bad instruction mix & Performance insensitive to problem size fitting into different cache levels & Large ratio of instructions retired to FP instructions if the useful work is FP / Many cycles per instruction (CPI) if the problem is large-latency arithmetic / Scalar instructions dominating in data-parallel loops (\verb+FLOPS_DP+, \verb+FLOPS_SP+, \verb+CPI+ is always measured)\\\hline
Limited instruction throughput & Large discrepancy between actual performance and simple predictions based on max Flop/s or LD/ST throughput & Low CPI near theoretical limit if instruction throughput is the problem / Static code analysis predicting large pressure on single execution port / High CPI due to bad pipelining (\verb+FLOPS_DP+, \verb+FLOPS_SP+, \verb+DATA+, \verb+CPI+ is always reported)\\\hline
Microarchitectural anomalies & Large discrepancy between actual performance and performance model & Relevant events are very hardware-specific, e.g., stalls due to 4k memory aliasing, conflict misses, unaligned vs. aligned LD/ST, requeue events. Code review required, with architectural features in mind.\\\hline
Synchronization overhead & Speedup going down as more cores are added / No speedup with small problem sizes / Cores busy but low FP performance & Large non-FP instruction count\footnotemark[2] (growing with number of cores used) / Low CPI (\verb+FLOPS_DP+, \verb+FLOPS_DP+, \verb+CPI+ always measured)\\\hline
False cache line sharing & Very low speedup or slowdown even with small core counts & Frequent (remote) evicts (\verb+CACHE+) \\\hline
Bad ccNUMA page placement & Bad/no scaling across locality domains & Unbalanced bandwidth on memory interfaces / High remote traffic (\verb+MEM+)\\
    \end{tabular}\normalsize\\[2mm]
    \caption{Performance patterns and corresponding signatures for parallel code on multicore systems}
    \label{tab:signatures}
\end{table}

\section{Case studies}\label{sec:cases}

\subsection{Abstraction penalties in C++ code}

The basis for this case study is a recent analysis of Expression
Template (ET) frameworks for basic linear algebra
operations~\cite{smartET12,hpcs12}. While classic ETs show good
performance for simple BLAS1-type (vector-vector) loop kernels, they
have severe problems with BLAS2- and BLAS3-type operations and sparse
arithmetic, since they are based on accesses to individual elements of
data structures and have no notion of standard optimizations for
nested loops. ``Smart Expression Templates'' (SETs) ameliorate
this problem since they provide a high-level approach to complex
loop nests and can substitute the whole operations by calls
to optimized libraries (such as Intel MKL) or well-written plain C 
or compiler intrinsics code. (S)ET approaches must also be compared
to standard coding techniques like operator overloading (which 
is plagued by the generation of temporary objects) and classic 
C loop nests. 

\footnotetext[2]{Load imbalance and frequent synchronization often go together, leading to large non-FP instruction counts that are caused by spin-waiting loops.}Convoluted code like the one generated by strongly abstract C++
source when dealing with matrix-type operands typically shows the ``\textbf{bad
instruction mix}'' pattern, since a lot of instructions are generated
that are not actually needed to solve the problem. In the Expression Template
example this shows most prominently in the number of retired instructions.
Table~\ref{tab:ET} shows events and derived metrics for a
5000$\times$5000 matrix-matrix multiplication using four different
code versions: The ``Classic'' code uses traditional overloading
of \verb.operator*(). so that an expression like \verb.C=A*B., with
\verb.A., \verb.B., and \verb.C. being objects of some matrix class,
results in a call to a function that implements a naive version of the
matrix multiply and returns the result as a temporary
copy. ``Boost \ublas'' supports matrix operations directly
with a slightly different syntax, and avoids the temporary (which is
the main reason for using ETs in the first place). \eigen\ is an SET
framework that is able to recognize arithmetic expressions involving
complex data types and employs an optimized version of the operation.
However, it still relies heavily on the inlining capabilities of the 
compiler. ``MKL dgemm'' denotes a direct call to the vendor-optimized
BLAS library for Intel processors.

The results show a striking agreement in the number of generated
instructions between the Classic and \ublas\ versions, although the
performance of the Classic code is a factor of eight higher. In both
cases the compiler has generated bloated, scalar machine code; the
reason is that the access to individual matrix elements is strongly
abstracted.  The ``Classic'' code, uses an overloaded
\verb.operator(int,int).  for accessing the matrix elements in the
loop nest, and the \ublas\ relies on a similar mechanism. Both
strategies impede the compiler's view on what the actual operation is
and limit its optimization capabilities. The result is a factor of
five to six in retired instructions compared to \eigen\ and MKL,
of which a factor of two can be attributed to scalar (as opposed to
SIMD-parallel) instruction code.

The fact that \ublas\ is so much slower than the Classic version
results from a very unfortunate loop ordering, leading to stride-5000
accesses to one of the matrices in the product (for details see
\cite{smartET12}). As a consequence, the code becomes latency-dominated
and makes inefficient use of the memory bandwidth (second column in 
Table~\ref{tab:ET}). This is the ``\textbf{strided data access}'' pattern at work.
The Classic version, despite its inefficient
machine code, at least implements a loop nest that has stride-one
accesses to all relevant data structures. 
This is also reflected in the CPI metric (fourth column), which indicates massive
pipeline bubbles due to long-latency loads.  The Classic
version can still not exhaust
the available memory bandwidth for a single thread (see ``STREAM''), although the
kernel should be bandwidth-bound. This is again a consequence of
the code spending too much time with in-core execution.

The \eigen\ version, with the help of optimized kernels and massive
inlining, achieves 76\% of the MKL performance, which is impressive
for compiler-generated code. The memory bandwidth of the MKL code
is not so different from the \ublas\ version, but this is purely
coincidental: Memory access is not a bottleneck for the highly
optimized dgemm implementation.

\begin{table}
\begin{center}
\begin{footnotesize}
\begin{tabular}{ll|p{2.1cm}|p{2.1cm}|p{2.1cm}|p{2.1cm}p{0.01cm}}
\multicolumn{2}{c|}{}   & \centering Memory Bandwidth [MByte/s] & \centering Total Retired Instructions [$10^{11}$] & \centering Cycles Per Instruction (CPI) & \centering Performance [MFlop/s] &\\[1mm]
\hline
             & STREAM              & \centering 11814                       & \centering ---                                    & \centering ---        & \centering --- &\\[1mm]
\hline
\multirow{6}{*}{\begin{sideways}\centering  $\quad$\end{sideways}} 
             & Classic             & \centering  5314    & \centering $12.5420$    & \centering $0.440861$ & \centering 1249 & \\[1mm]
             & Boost \ublas        & \centering   630    & \centering $10.1207$    & \centering $4.61834$  & \centering 156  & \\[1mm]
             & \eigen              & \centering   371    & \centering $ 2.1014$    & \centering $0.41168$  & \centering 8555 & \\[1mm]
             & MKL dgemm           & \centering   531    & \centering $ 2.03448$   & \centering $0.321115$ & \centering 11261 & \\[1mm]
\end{tabular}
\end{footnotesize}
\end{center}
\caption[Hardware counter performance analysis of the multiplication of two dense matrices]{
   \label{tab:ET}
   Hardware counter performance analysis of the single-threaded 
   multiplication of two large dense matrices 
   ($N = 5000$). The given STREAM bandwidth is the practical limit for
   one thread on the used processor. (Adapted from \cite{smartET12})
}
\end{table}

\subsection{Medical image reconstruction by backprojection}

\begin{figure}[tb]\centering
    \subfloat[FP performance]{\includegraphics*[height=0.67\linewidth]{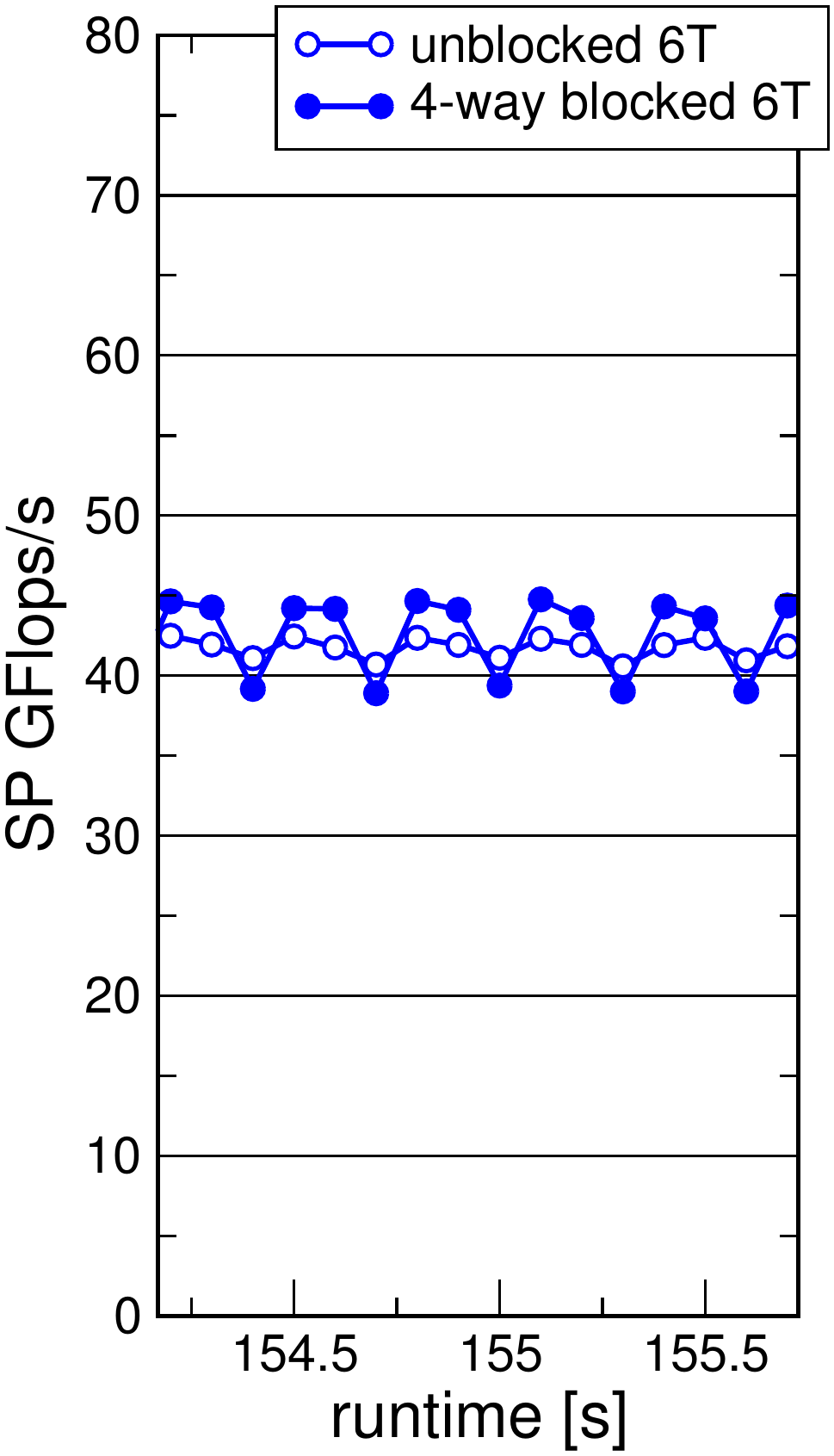}}\hfill
    \subfloat[Memory bandwidth]{\includegraphics*[height=0.67\linewidth]{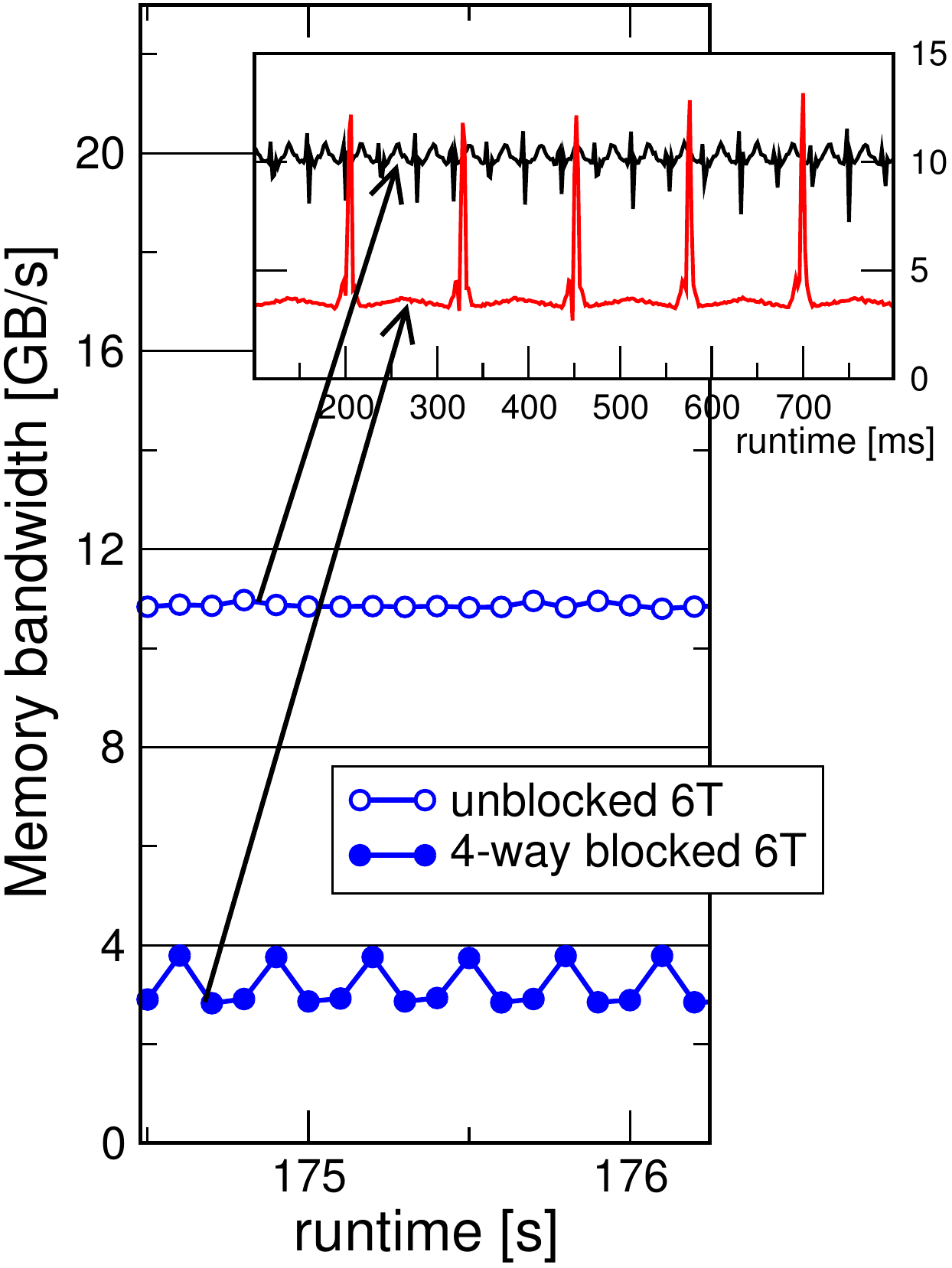}}
    \caption{Performance counter timeline monitoring of floating-point performance (a)
	and  memory bandwidth (b), comparing blocked/nonblocked 
	variants of the best implementation on one full Westmere socket (six cores)
        at 100\,ms resolution. 
	The inset in (b) shows a zoomed-in view with 2\,ms resolution. 
        (Adapted from \cite{rabbitct})}
    \label{fig:perfctr}
\end{figure}

This case study was part of a work aiming at optimized volume reconstruction
on multicore processors~\cite{rabbitct}. The
optimization target is the open benchmark RabbitCt, which implements volume
reconstruction by backprojection, which is the computational
bottleneck in many medical imaging applications. From a simple roofline model
analysis \cite{Williams:EECS-2008-134} the algorithm was identified to be 
bandwidth-limited on the platforms
investigated. However, it turned out that a complex combination of patterns
is involved here: \textbf{memory bandwidth saturation}, \textbf{limited instruction throughput},
and \textbf{load imbalance}. 

In a first attempt it was checked if the memory bandwidth saturation
pattern applies using the \verb+MEM+ performance group of likwid-perfctr. 
To get a meaningful bandwidth baseline a benchmark was constructed that
mimics the basic data access pattern, which in this case is an array 
update kernel (\verb+A(:)=s*A(:)+). 
The Intel Westmere processor used in the analysis 
can sustain 20.3\~GB/s using all cores of one socket for this
operation. Bandwidth measurements with likwid-perfctr revealed that the
application showed a much lower bandwidth of roughly 10\~GB/s. Hence,
memory bandwidth saturation is not a limiting bottleneck for this
implementation on Intel Westmere. 
A static code analysis (using the Intel IACA tool \cite{iacaweb})
showed that the scattered load of necessary pixel data into SIMD registers
requires a large number of instructions on the instruction code level.  This
makes the code limited by instruction throughput, which is
not evident from the high-level language implementation. The measured performance was
near the prediction of this static loop body runtime analysis, which was based
on the instruction throughput capabilities of the architecture.  A final
confirmation that the code is indeed limited by instruction throughput was
achieved by comparing with measured CPI values, which were in accordance with the
static L1 cache prediction.

For a more severely bandwidth-starved architecture (Intel Harpertown) a
cache-blocked version of RabbitCt was implemented, showing a significant
performance improvement. This code version was also run on the
Westmere platform for comparison. The result of a likwid-perfctr timeline measurement
of the floating-point performance and main memory bandwidth is shown in 
Fig.~\ref{fig:perfctr}: The blocking effectively lowers the bandwidth
demands, but there is no impact on the overall performance.

One of the other applied optimizations was a work reduction strategy;  after
cutting over 30\% off the total work the runtime benefit was still negligible. To
check a possible load imbalance the instruction count on the cores was measured
using likwid-perfctr with the \verb+FLOPS_SP+ group. As the innermost loop body
is fully vectorized the number of packed (vectorized) arithmetic instructions
is a good indicator for a potential load imbalance. Table \ref{tab:loadi} shows the
results of a likwid-perfctr measurement of the  packed SSE arithmetic floating 
point instructions.
\begin{table}[tb]
\begin{center}
    \begin{tabular}{c|c|c|c|c|c|c}
        Core Id&0&1&2&3&4&5\\
        \hline
    \verb+FP_COMP_OPS_EXE_SSE_FP_PACKED+  [$10^{10}$]& 2.74 & 9.39 & 9.23 & 9.30 & 9.29  & 3.07
    \end{tabular}
\end{center}
    \caption{Instruction count per core for packed SSE arithmetic floating point instructions
        of the RabbitCt benchmark without load balancing}
    \label{tab:loadi}
\end{table}
Evidently the outer threads have only one third of the workload of the others 
in terms of packed FP instructions. The runtime for this case is 61.72\,s.  
A simple way to improve load balancing in in OpenMP was to change the loop 
scheduling to a round robin distribution, using \verb+static,1+. 
Results are shown in Table \ref{tab:loadb}: The load imbalance was 
removed and the runtime was reduced to 43.9\,s.
\begin{table}[bt]
\begin{center}
    \begin{tabular}{c|c|c|c|c|c|c}
        Core Id&0&1&2&3&4&5\\
        \hline
    \verb+FP_COMP_OPS_EXE_SSE_FP_PACKED+  [$10^{10}$]& 7.16 & 7.17 & 7.16 & 7.17 & 7.17  &7.17 
    \end{tabular}
\end{center}
\caption{Instruction count per core for packed SSE arithmetic floating point instructions with round robin scheduling}
    \label{tab:loadb}
\end{table}

\section{Summary and outlook}\label{sec:summary}

This paper presented an initial proposal of a structured usage of HPM embedded
in an overall software engineering process. We classify
relevant performance patterns and formulate signatures which indicate that a
certain pattern applies. The signatures are based on HPM data alone or combined
with other sources of information such as microbenchmarking data and static code and
algorithmic analysis. We are aware that this approach nevertheless requires an
intimate knowledge of the algorithm, the code and the hardware. Still we
believe that there is no alternative to a performance engineering process build
on knowledge of the programmer himself. The application of our approach was
illustrated on the example of two case studies. Future work involves further
settlement of the performance patterns and corresponding signatures.
This will be achieved by applying the approach to various practical examples.

\bibliographystyle{drgh}
\bibliography{rrze}

\end{document}